\documentclass[prb,aps,showpacs,superscriptaddress,floatfix,twocolumn]{revtex4-1}

\usepackage{graphicx}
\usepackage{dcolumn}
\usepackage{amsfonts,amsmath,amssymb,bm}

\begin{document}

\makeatother
\newcommand*{\rom}[1]{\expandafter\@slowromancap\romannumeral #1@}
\makeatother

\title{Ab initio analysis of some Ge-based 2D nanomaterials}
\date{\today}
\author{Ali Ghojavand}
\email{a.ghojavand@ph.iut.ac.ir}
\affiliation{Department of Physics, Isfahan University of Technology,
Isfahan, 84156-83111, Iran}
\author{S. Javad Hashemifar}
\email{hashemifar@iut.ac.ir}
\affiliation{Department of Physics, Isfahan University of Technology,
Isfahan, 84156-83111, Iran}
\author{Mahdi Tarighi Ahmadpour}
\affiliation{Department of Physics, Isfahan University of Technology,
Isfahan, 84156-83111, Iran}
\author{Alexander V. Shapeev}
\affiliation{Skolkovo Institute of Science and Technology,
Skolkovo Innovation Center, Bolshoy Boulevard 30, bld.\ 1, Moscow, 121205, Russia}
\author{Amir Alhaji}
\affiliation{Department of Materials Engineering, Isfahan University of Technology,
Isfahan 84156-83111, Iran}
\author{Qaem Hassanzada}
\affiliation{Department of Physics, Isfahan University of Technology,
Isfahan, 84156-83111, Iran}

\begin{abstract}

The structural, electronic and dynamical properties of a group
of 2D germanium-based compounds,
including GeC, GeN, GeO, GeSi, GeS, GeSe, and germanene, 
are investigated by employing first-principles calculations. 
The most stable structure of each of these systems is identified after 
considering the most probable configurations and performing
accurate phonon calculations.
We introduce a new phase of germanene, which we name the tile germanene,
which is significantly more stable than the known hexagonal germanene.
We apply the modern modified Becke-Johnson (mBJ) and DFT1/2 schemes
to obtain an accurate band structure for our selected 2D materials.
It is seen that GeO and GeC exhibit the highest band gaps of 
more than 3\,eV in this group of materials.
Moreover, we argue that, in contrast to the semi-metallic nature of hexagonal germanene,
the tile germanene is a very good conductor.
The band edges of our semiconducting 2D materials are accurately aligned to
the vacuum level to address the potential photocatalytic application of this system for
water splitting and carbon dioxide reduction.
The optical properties, including dielectric functions,
refractive index, reflectivity, and Loss function of the samples
are investigated in the framework of the Bethe-Salpeter approach.

\end{abstract}

\maketitle

\section{INTRODUCTION}

In recent years, elemental sheet of germanium, known as germanene, 
has been emerging as a strong contender in the realm of 2D materials. 
Germanene is a one-atom-thick germanium layer which has 
a honeycomb structure (D3d point group) and a zero band gap 
with a Dirac cone at the K point of the Brillouin zone. 
In 2009, Cahangirov et al, by using first-principles calculations,
predicted a low buckled (corrugated) sheet structure for 
a 2D germanene layer.\cite{cahangirov2009}
The main hurdle experienced in realizing individual 
germanene layers is that, unlike graphene, 
they do not form a van der Waals layered structure in their natural form. 
Hence, top-down approaches are not applicable for
synthesis of individual germanene layers.
In 2014, the first synthesis of germanene was realized on a gold (111) substrate
with a growth mechanism similar to the formation of silicene layers
on silver (111) templates.\cite{Davila2014}

The absence of band gap dramatically hampers direct applications of germanene
in semiconductor devices in nanoelectronics, photoelectronics, and sensors. 
Hence, seeking an effective method to open a sizable band gap in germanene 
is an active field of research. 
Chemical functionlization with small molecules,
introducing structural defects, and alloying with proper elements
are three conventional methods to engineer the band structure of 2D materials.
Hydrogen functionalized germanene layers (GeH) were successfully synthesized
in 2013 with a band gap of about 1.5\,eV and a similar structure to graphane.\cite{geh2013}
Padilha and others considered Stone-Wales (SW), single vacancy, and divacancies defects
in germanene and showed that the SW defect open a band gap in the system
and destroys the Dirac cone, while the single vacancy defect preserves the Dirac cone.\cite{padilha2016}
Introducing structural defects such as Stone–Wales, 
single vacancy, and divacancies strongly affects the band structure and 
transport properties of the system, compared with the pristine one.
Xu et al. predicted that alloying with Se may lead to the formation of 
two different semiconducting configurations of GeSe monolayers,
which exhibit anisotropic absorption spectra in the visible region.\cite{xu2017}

\begin{figure}
 \includegraphics[scale=0.25]{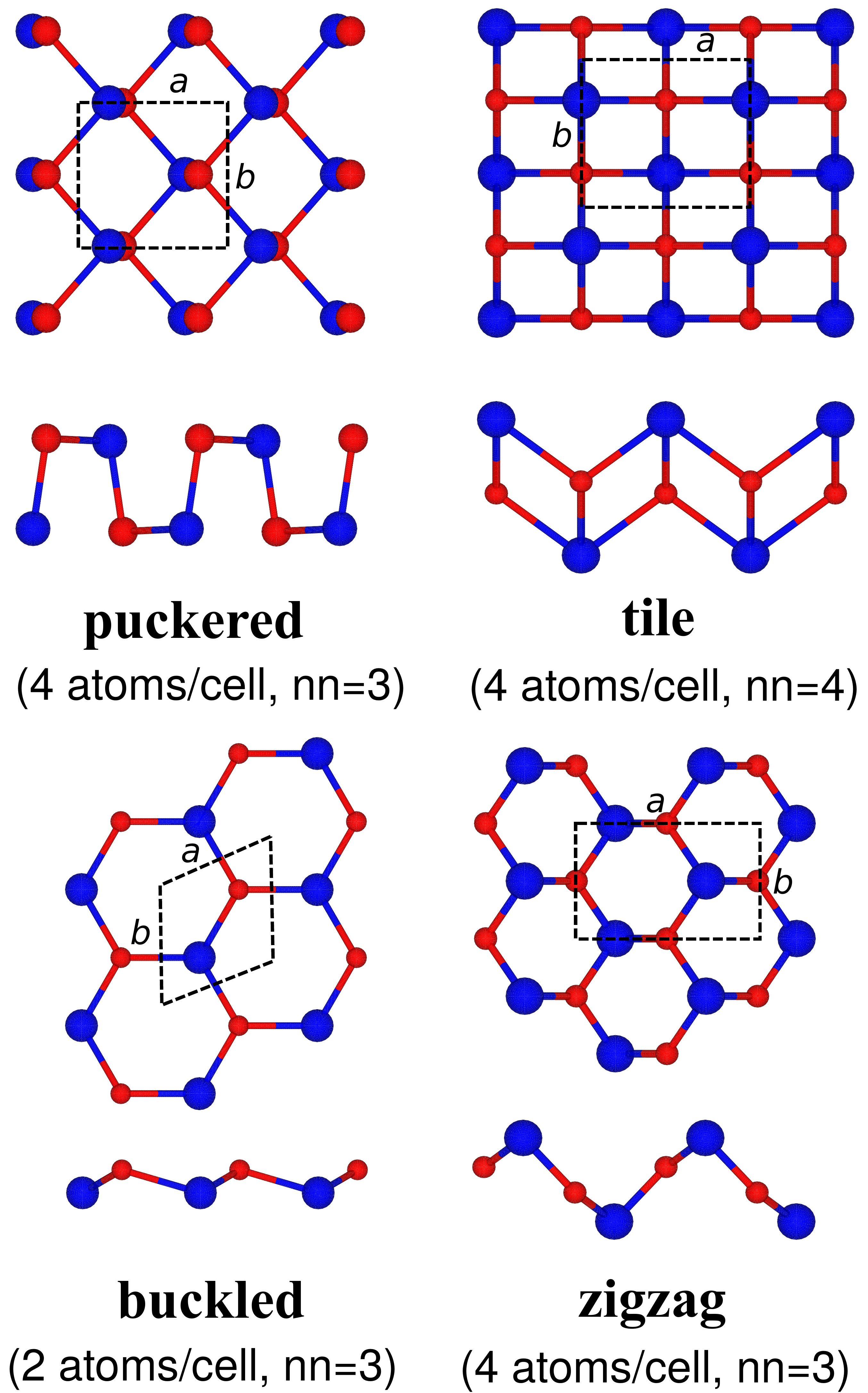}
 \caption{\label{structure}
 Top and side views of the four candidate structures for
 our Ge based 2D materials.
 The dotted lines encapsulate the 2D unitcell of the lattices
 and the blue and red balls indicate two nonequivalent atomic positions
 in the systems.
}
\end{figure}

In this work, we employ first-principles calculations to study 
the structural, electronic, and optical properties and dynamical stability 
of a group of Ge-based 2D materials including GeO, GeS, GeSe, GeN, GeC, GeSi, and germanene.
In the next section, we introduce the computational techniques used in this work.
Then the stability of the systems will be addressed in section III,
electronic properties will be explained in section IV,
and optical properties will be presented in section V.
The summary of our work will be given in the last section.

\section{METHOD}

We performed electronic structure calculations and geometry optimization 
in the framework of density functional theory (DFT)\cite{hohenberg1964,kohn1965} and 
the Perdew, Burke, and Ernzerhof (PBE) exchange–correlation functional\cite{perdew1996} 
by using the full-potential all-electron numeric atom-centered orbital method 
implemented in the FHI-aims package.\cite{blum2009ab}
In order to obtain very accurate binding energies and optimized geometries,
the structure relaxation tolerance was set to 0.001 eV/\AA\.
In order to verify dynamical stability of the structures,
phonon calculations were performed by using the density functional
perturbation theory and the plane wave ultrasoft pseudopotential method,
implemented in the QUANTUM ESPRESSO package.\cite{giannozzi2009quantum}
The reliability of the pseudopotentials was verified by comparing 
the obtained binding energies in the Quantum Espresso and FHI-aims packages.
The novel DFT1/2 and modified Becke-Johnson (mBJ) schemes 
were applied to correct the PBE electronic band structure of the systems.
The accuracy of the band gaps in these methods and especially DFT1/2 were  
comparable with the expensive hybrid functional and GW methods,\cite{hedin1965,godby1989,hybertsen1986}
while their required computational effort is comparable to that of PBE.
The DFT1/2 method\cite{ferreira2008approximation}, which extends the Slater’s half-occupation
technique to bulk materials, were applied by using the Exciting package 
which employs the full potential linear augmented plane wave method
to solve the single particle Kohn-Sham equations.\cite{gulans2014exciting}
On the other hand, the modified Becke-Johnson (mBJ) method\cite{tran2009accurate}
was applied by using the Wien2k package\cite{wien2k} which 
has a very similar technical structure to the Exciting package.
The monolayer structures were simulated in the slab supercells 
with a vacuum thickness of about 13\AA\ 
to avoid unrealistic effects from periodic boundary conditions.

The optical properties of the systems were calculated by using the Exciting package
which takes into account the nontrivial effects of electron-hole interaction and 
solves the Bethe-Salpeter equation\cite{onida2002} (BSE):

\[(\varepsilon_{c\mathbf k}-\varepsilon_{v\mathbf k})A_{cv\mathbf k}^S+\sum_{\mathbf k' c' v'}
\kappa_{c'v'\mathbf k'}^{cv\mathbf k}A_{c'v'\mathbf k'}^S=\Omega^sA_{cv\mathbf k}^{S}\]
where the term ($\varepsilon_{c\mathbf k}-\varepsilon_{v\mathbf k}$) refers to the difference 
between the conduction and valence quasiparticle energies
at a specific k-point, $\kappa$ describes the electron-hole interaction,
and $\Omega^{s}$ is the excitation energy.
After solving for the BSE excitation states, the Tomm-Dancoff approximation\cite{rohlfing2000,benedict1998,van1999,fetter2012} (TDA) 
is used to compute the imaginary part of the dielectric function ($\varepsilon_2$):

\[{\rm Im} (\varepsilon_{M}^i(\omega))=\varepsilon_2(\omega)=
  \frac{16\pi^2e^2}{\omega^2}\sum_{S}|\vec e.\langle0|\vec v|S\rangle|^2\delta(w-\Omega^s)\]
where $\vec e$ describes the polarization of the incident
light and $\vec v$ is the velocity operator.
Then the Kramers-Kronig relations are employed to find 
the real part of the dielectric function ($\varepsilon_1$) and 
subsequently other linear optical properties including
refractive index $n(\omega)$, reflectivity $R(\omega)$,
and electron loss function $L(\omega)$:

\[n(\omega)=\left(\frac{\sqrt{\varepsilon_1^2+\varepsilon_2^2}+\varepsilon_1}{2}\right)^{1/2}\] 
\[k(\omega)=\left(\frac{\sqrt{\varepsilon_1^2+\varepsilon_2^2}-\varepsilon_1}{2}\right)^{1/2}\] 
\[R(\omega)=\frac{(n-1)^2+k^2}{(n+1)^2+k^2}\hspace{1cm}  
L(\omega)=\frac{\varepsilon_2}{\varepsilon_1^2+\varepsilon_2^2}\]

\section{Stable structures}

After a broad literature survey, we realized that the structures of 
the most novel 2D materials may be generally categorized in the puckered 
and buckled configurations, presented in Fig.~\ref{structure},
while other structures are rarely investigated in the literature.
Hence, we applied these 2D structure patterns to our desired materials;
Ge, GeO, GeS, GeSe, GeN, GeC, and GeSi.
During geometrical optimizations, in some cases, we noticed appearance of 
two new structures, called zigzag and tile in Fig.~\ref{structure},
which were added to the candidate structures of our 2D materials.
In order to find the most stable structure of each of the systems, 
we calculated their binding energy in the four candidate structures 
by comparing the optimized total energies with 
the energy of corresponding isolated atoms.
The obtained binding energies are presented in table~\ref{binding}.

\begin{table}
\caption{\label{binding}
 Calculated binding energy (eV/atom) and thickness (\AA) (in the parenthesis)
 of the studied 2D materials in the four candidate structures.}
\begin{ruledtabular}
\begin{tabular}{lcccc}
     &    buckled       &   puckered       &      zigzag      &   tile           \\
\hline                                    
GeO  & $-$5.60 (0.99)     &     ---          &{\bf$-$5.85} (2.23) &    ---           \\
GeS  & $-$4.48 (1.36)     &{\bf$-$4.51} (2.57) &       ---        &    ---           \\
GeSe & $-$4.16 (1.44)     &{\bf$-$4.17} (2.60) &       ---        &    ---           \\
GeN  & $-$5.66 (0.00)     &    ---           &{\bf$-$6.01} (1.26) &    ---           \\
GeC  &{\bf$-$5.98} (0.00) &    ---           &       ---        &    ---           \\
GeSi & $-$4.34 (0.58)     &    ---           &       ---        &{\bf$-$4.52} (1.95) \\
Ge   & $-$4.01 (0.67)     &     ---          &       ---        &{\bf$-$4.12} (1.97) \\
\end{tabular}
\end{ruledtabular}
\end{table}

\begin{table}
\caption{\label{lattice}
  Relaxed structural parameters \textit{a,b} and 
  average nearest neighbor bond lengths d1 and d2 (\AA) of the
  investigated 2D systems, in their most stable structure.
}
\newcommand{\guo}{\cite{guo2019}}
\newcommand{\gom}{\cite{gomes2015}}
\newcommand{\xu}{\cite{xu2017}}
\newcommand{\sah}{\cite{sahin2009}}
\begin{ruledtabular}
\begin{tabular}{cccccccc}
    & GeO  & GeS  & GeSe &  GeN &  GeC & GeSi & Ge-tile \\
\hline                                      
$a$ & 4.64 & 4.54 & 4.29 & 5.36 & 3.23 & 3.94 & 4.19 \\
$b$ & 3.01 & 3.63 & 3.98 & 3.06 & 3.23 & 3.64 & 3.83 \\
d1  & 1.97 & 2.43 & 2.54 & 2.17 & 1.87 & 2.69 & 2.71 \\
d2  & 1.97 & 2.46 & 2.66 & 1.96 & 1.87 & 2.58 & 2.71 \\
\end{tabular}
\end{ruledtabular}
\end{table}

\begin{figure*}
\includegraphics[scale=0.9]{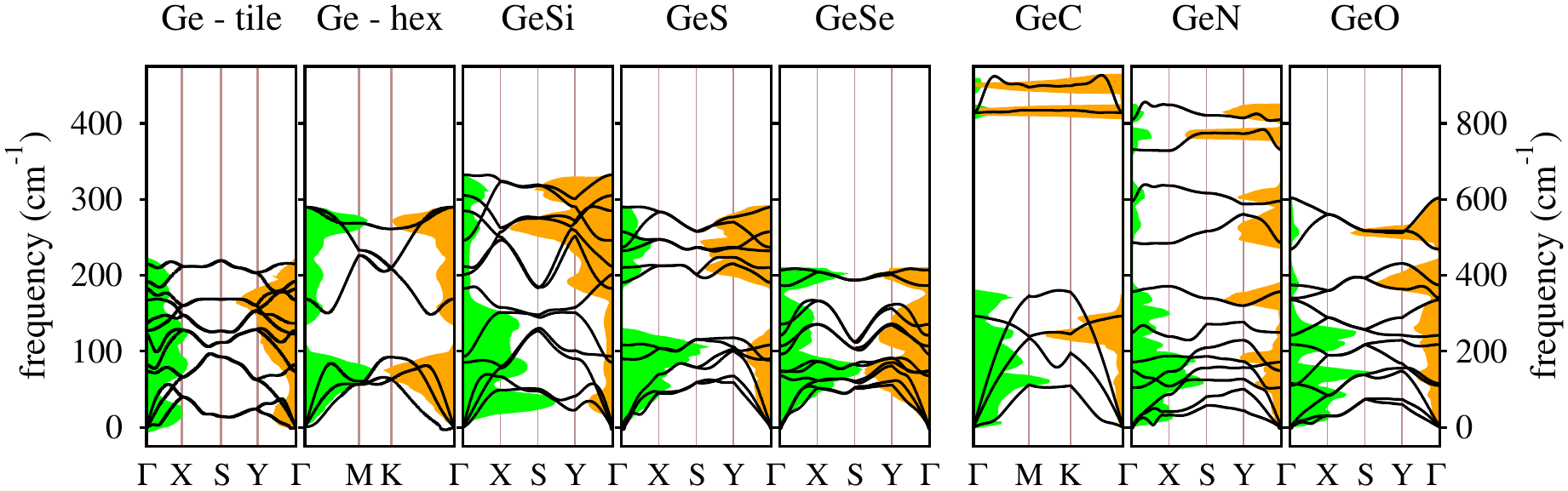}
\caption{\label{phonon}
 Obtained phonon band structure of our selected 2D materials. 
 The green and orange shaded areas show 
 the phonon partial density of states of Ge and its partner atoms
 in our 2D systems, respectively.
}
\end{figure*}

\begin{figure}
 \includegraphics[scale=0.9]{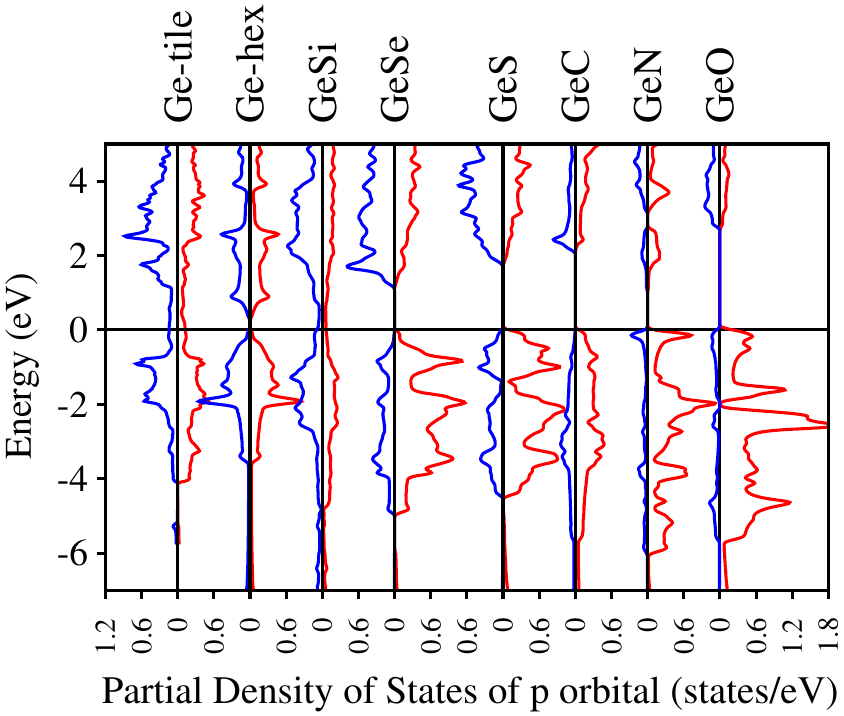}
 \caption{\label{dos}
 Contribution of p orbitals of Ge (blue, left oriented) 
 and its partner atom (red, right oriented)
 in the density of states of the investigated 2D systems.
}
\end{figure}

It is very interesting to see that these materials mostly admit 
the buckled configuration as a metastable structure while
their lowest-energy structure is among other candidates.
It is more fascinating in the case of germanene, 
where its well known hexagonal structure (zero buckled) is 
considerably less stable than our discovered tile structure.
A more reliable and detailed investigation of this issue 
requires studying dynamical stability which 
will be presented later.
The obtained results indicate that the binding energy is decreasing 
with the increase of the atomic radius. 
In other words, oxygen, nitrogen, and carbon atoms which
are the smallest atoms in our samples give rise to the highest
binding energies for the GeO, GeN, and GeC compounds.
On the other hand, Ge which has the largest atomic radius
leads to the lowest binding energy for the 2D germanene.
This trend indicates a stronger bonding between atoms 
with smaller radii which is in agreement with the physical intuition.
The equilibrium lattice constants and bond lengths of 
the lowest-energy structures of GeO, GeS, GeSe, GeN, GeC, GeSi, 
and germanene are calculated and presented in Table~\ref{lattice}.

As it was mentioned before, phonon calculations should be done
to confirm the stability of the lowest-energy structures.
The phonon dispersion curves of all the materials at their lowest-energy structure 
were computed by using the density functional perturbation theory method. 
The obtained phonon spectra of the samples along their high symmetry
paths in the reciprocal space are presented in Fig.~\ref{phonon}.
The absence of any negative (imaginary) frequency mode in the spectra
demonstrates dynamical stability of all the systems.
We observe that GeC, GeN, and GeO display the widest range of phonon modes 
among the studied systems, providing further evidence for stronger
bonding in these systems, compared with others.
Especially in GeC, a large distance seen between the acoustic modes and 
the two optical modes which is an evidence of a strong directional bonding in this system.
In the new phase of germanene, in contrast to the hexagonal phase,
we observe that the optical modes are not well separated from the acoustic modes,
which may be attributed to the softer bonding in the tile germanene.

For better understanding of the dynamical features of the systems,
we calculated the phonon partial density of states (PDOS) of Ge and its partner atoms
in our investigated 2D materials.
The results are presented as shaded areas in Fig.~\ref{phonon}.
It is clearly seen that the heavy elements Ge and Se have a larger
contribution to the acoustic phonon modes while the light elements
exhibit stronger vibrations in the optical modes.

Dynamical stability of the new phase of germanene along with its lower binding energy,
compared with the hexagonal germanene, raises the question of why this phase has not yet
been observed in real samples.
In order to address this question one should note that within all successful 
synthesis of germanene, the [111] surface of gold or platinum has been used as the substrate.
These surfaces involve hexagonal arrangement of the substrate atoms which
may enhance formation of hexagonal germanene on the surface.
Hence, atomistic growth of germanene atoms on square-symmetry surfaces may 
enhance the formation of tile germanene in realistic samples.

\section{Electronic structure}

The obtained electronic PDOS of the systems in their 
lowest-energy structure, within PBE, are presented in Fig.~\ref{dos}.
It is seen that the systems with the highest binding energies
(GeO, GeN, and GeC) exhibit the most broadened valence bands,
in agreement with strong bonding in these systems.
Moreover, the valence p orbital of Ge in GeO and GeN
is effectively evacuated and transferred to 
the valence p orbital of its partner atom,
indicating significant ionic bonding in these two monolayers.
GeSi and the tow configurations of germanene represent a metallic 
electronic structure while the other investigated 2D materials
are semiconductors.
The tile configuration of germanene is likely a very good conductor,
because it has a finite density of high mobility p electrons at the Fermi level.

\begin{figure}
   \includegraphics[scale=0.65]{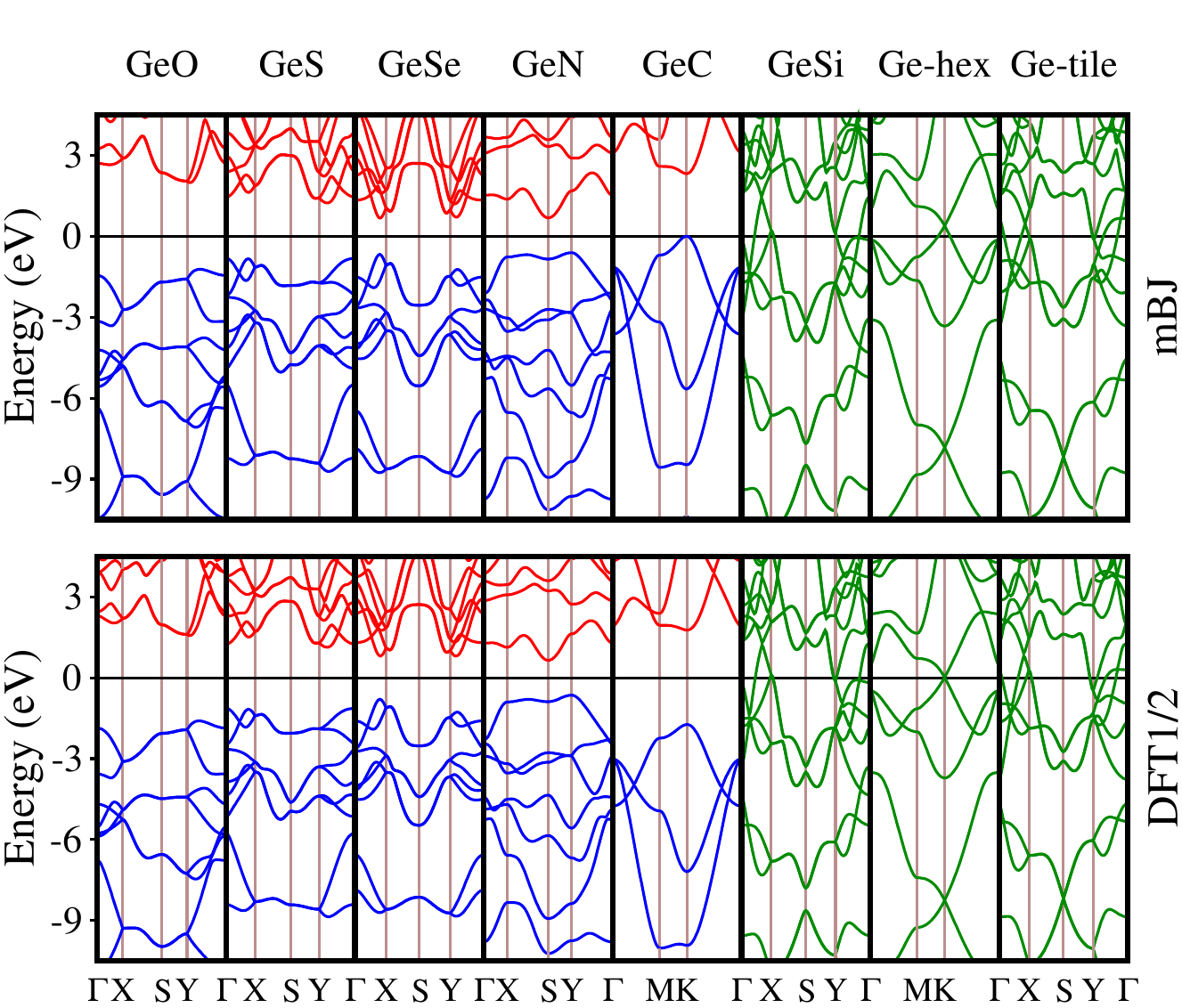}
   \caption{\label{band}
   Electronic band structures of our Ge based 2D compounds 
   within the mBJ (top row) and DFT1/2 (bottom row) methods.
}
\end{figure}

In order to obtain more accurate electronic structures, 
as it was mentioned in the Methods,
we apply the DFT1/2 and mBJ schemes which have been proven 
to predict much more accurate band gaps,
compared with the conventional LDA and GGA functionals.
The calculated electronic band structures of GeO, GeS, GeSe, GeN, GeC and GeSi 
within the DFT1/2 and mBJ schemes along the high-symmetry directions of the Brillouin zone
are presented in Fig.~\ref{band}.
It is seen that these two methods predict quite similar band structures,
except for the position of Fermi levels which are occasionally different.
More discussion on the relative positions of Fermi levels and band edges
needs to the alignment of the energy references which will be presented 
in the next paragraphs.
Among the studied systems, GeC exhibit the most peculiar band structure
with very high valence band dispersion,
indicating a very stiff directional bonding in this system.
This observation is consistent with the large distance observed
between the acoustic and optical phonon modes of this system (Fig.~\ref{phonon}).
We observe that the Dirac cone of hexagonal germanene is slightly
shifted within mBJ, indicating less accuracy of the mBJ Fermi level, 
compared with DFT1/2.
GeSi and tile germanene exhibit similar band structures,
because both systems stabilize in the tile structure and have
similar atomic valence shell.
The same similarity is visible between GeS and GeSe band structures.
The calculated values for 
the energy gaps are summarized in Table \ref{bandgap}. 
In this table the reported band gap within the hybrid HSE06 functionals\cite{heyd2003,ge2006}
are also given for comparison. 
We observe that the predicted band gaps within DFT1/2 are closer to 
the HSE06 gaps, compared with the mBJ functional.
Due to the lack of experimental data, we are not able to compare 
the accuracy of the DFT1/2 and HSE06 methods,
although in the case of bulk semiconductors there is some
evidences for a higher accuracy of the DFT1/2 scheme.\cite{doumont2019}

\begin{table}
\caption{\label{bandgap}
  Calculated band gap (eV) of GeS, GeSe, GeC, GeO, and GeN
  in their most stable structures within the PBE, DFT1/2 and mBJ methods. 
  The most accurate reported band gaps within HSE06 and GW0 scheme
  are also given in the last column as the best theoretical references.
}
\newcommand{\guo}{\cite{Guo2019}}
\newcommand{\gom}{\cite{Gomes2015}}
\newcommand{\xu}{\cite{xu2017}}
\newcommand{\sah}{\cite{sahin2009}}
\begin{ruledtabular}
\begin{tabular}{lcccc}
      &   PBE  & DFT1/2 &  mBJ  &   others    \\
\hline                                             
GeO   &  2.84  & 3.22   & 3.27  &  3.73 [HSE]\guo \\
GeS   &  1.79  & 2.27   & 2.06  &  2.32 [HSE]\gom \\
GeSe  &  1.22  & 1.68   & 1.41  &  1.61 [HSE]\xu~~\\
GeN   &  1.05  & 1.29   & 1.30  &  ---            \\
GeC   &  2.06  & 3.50   & 2.32  &  3.56 [GW]\sah~ \\
\end{tabular}
\end{ruledtabular}
\end{table}

2D materials are generally considered as potential candidates for 
photocatalytic applications in various chemical reactions. 
Because in these nanomaterials, the relative surface area is very large,
the transport distance for the photo generated carriers to reach 
the reaction interface is very short,
and the band gap is likely enlarged due to the quantum confinement effect\cite{hu2019}.
Therefore, we screen the band edges of our germanium-based 2D materials 
to investigate their potential photocatalytic application in 
the water splitting and carbon dioxide conversion reactions.
The water splitting reactions are:
\vspace{2mm}\newline
2H$_2$O + 4h$^+$ \(\rightarrow\) ~4H$^+$ + O$_2$\hspace{2.6cm}($-5.67$\,eV) 
\vspace{2mm}\newline
2H$^+$ + 2e$^-$ \(\rightarrow\) ~H$_2$\hspace{4cm}($-4.44$\,eV) 
\vspace{2mm}\newline
where h$^+$ and e$^-$ are the photo-generated hole and electron
and numbers in the parenthesis are the corresponding redox potentials 
at room temperature and zero pH\cite{trasatti1986}.
A proper photocatalyst for these reactions should have
a conduction band bottom (CBB) below the hydrogen evolution potential
and a valence band top (VBT) above the oxygen evolution potential.
In addition to the above water splitting reactions,
we considered carbon dioxide conversion to methanol, 
formic acid, and methane as follows: 
\vspace{0mm}\newline
CO$_2$ + 6H$^+$ + 6e$^-$ \(\rightarrow\) ~CH$_3$OH + H$_2$O\hspace{0.9cm}($-4.06$\,eV)
\vspace{2mm}\newline
CO$_2$ + 2H$^+$ + 2e$^-$ \(\rightarrow\) ~HCOOH\hspace{2.0cm}($-3.83$\,eV)
\vspace{2mm}\newline
CO$_2$ + 4H$^+$ + 8e$^-$ \(\rightarrow\) ~CH$_4$ + H$_2$O\hspace{1.4cm}($-4.20$\,eV)
\vspace{2mm}\newline
As it was mentioned before, the numbers in the parenthesis are 
corresponding reduction potentials at room temperature and zero pH\cite{kanan2012}.
In order to consider other temperature and pH values, 
the redox potentials should be shifted by pH$\times$(K$_B$T$\times\ln10$).
The CBB of the proper semiconductor photocatalyst for the 
above mentioned conversions should be above 
the corresponding reduction potentials. 

\begin{figure}
 \includegraphics*[scale=0.85]{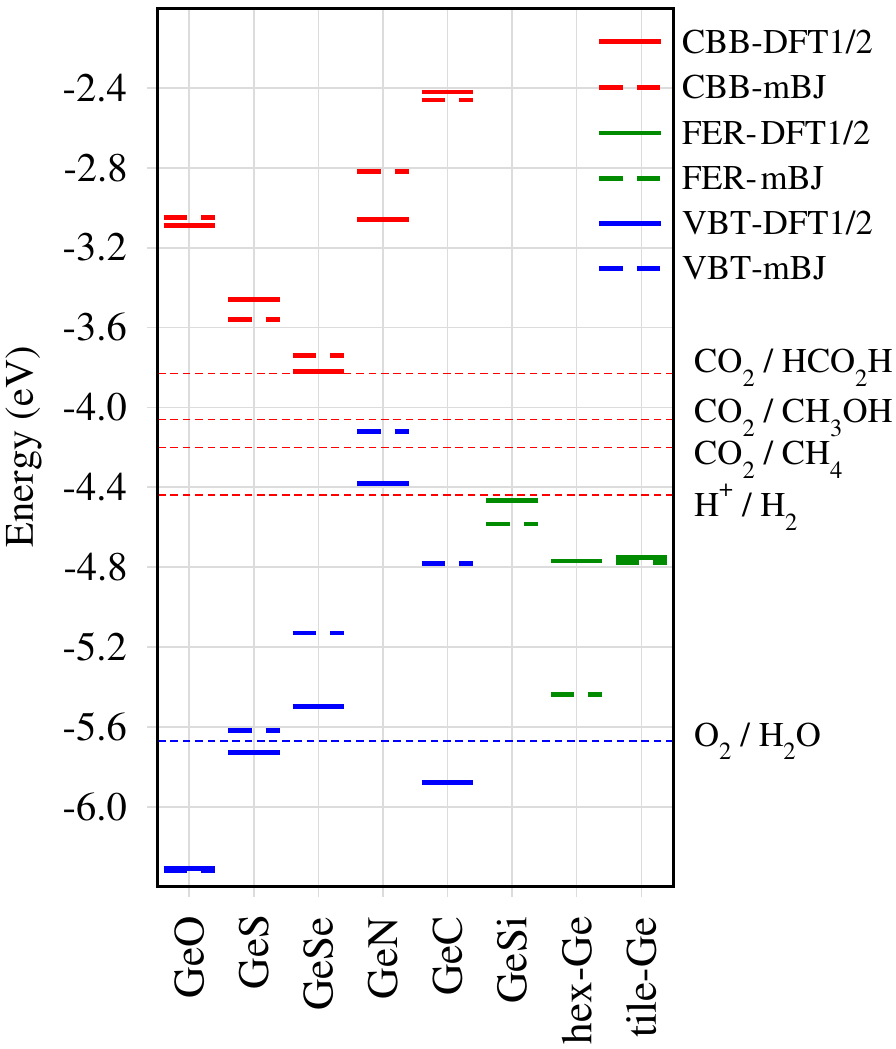}
 \caption{\label{catalyst}
 Valence band top (VBT), conduction band bottom (CBB), and
 Fermi (FER) levels of the studied 2D systems with respect to the vacuum level
 within DFT1/2 and mBJ schemes.
}
\end{figure}

The band edges of our 2D systems were determined with respect 
to the vacuum level potential within the DFT1/2 and mBJ schemes
and the resulting CBBs, VBTs, and Fermi levels 
are compared with the above mentioned redox potentials in Fig.~\ref{catalyst}.
The vacuum level is determined by averaging the electrostatic potential 
of the slab supercells in the horizontal planes and then plotting
the averaged potential as a function of vertical position z.
The obtained CBB band edges clearly indicate that all
our five 2D semiconductors (GeO, GeS, GeSe, GeN, and GeC)
are good photocatalyst candidates for the three considered 
carbon dioxide reduction reactions and the H$^+$/H$_2$ water splitting
half-reaction.
On the other hand, GeO, GeS, and GeC may be good photocatalysts for the 
H$_2$O/O$_2$ water splitting half-reaction 
while GeSe needs a rather little bias of less than 0.2\,eV
to photocatalyse this reaction.

\section{Optical properties}

\begin{figure*}
\includegraphics[scale=0.90]{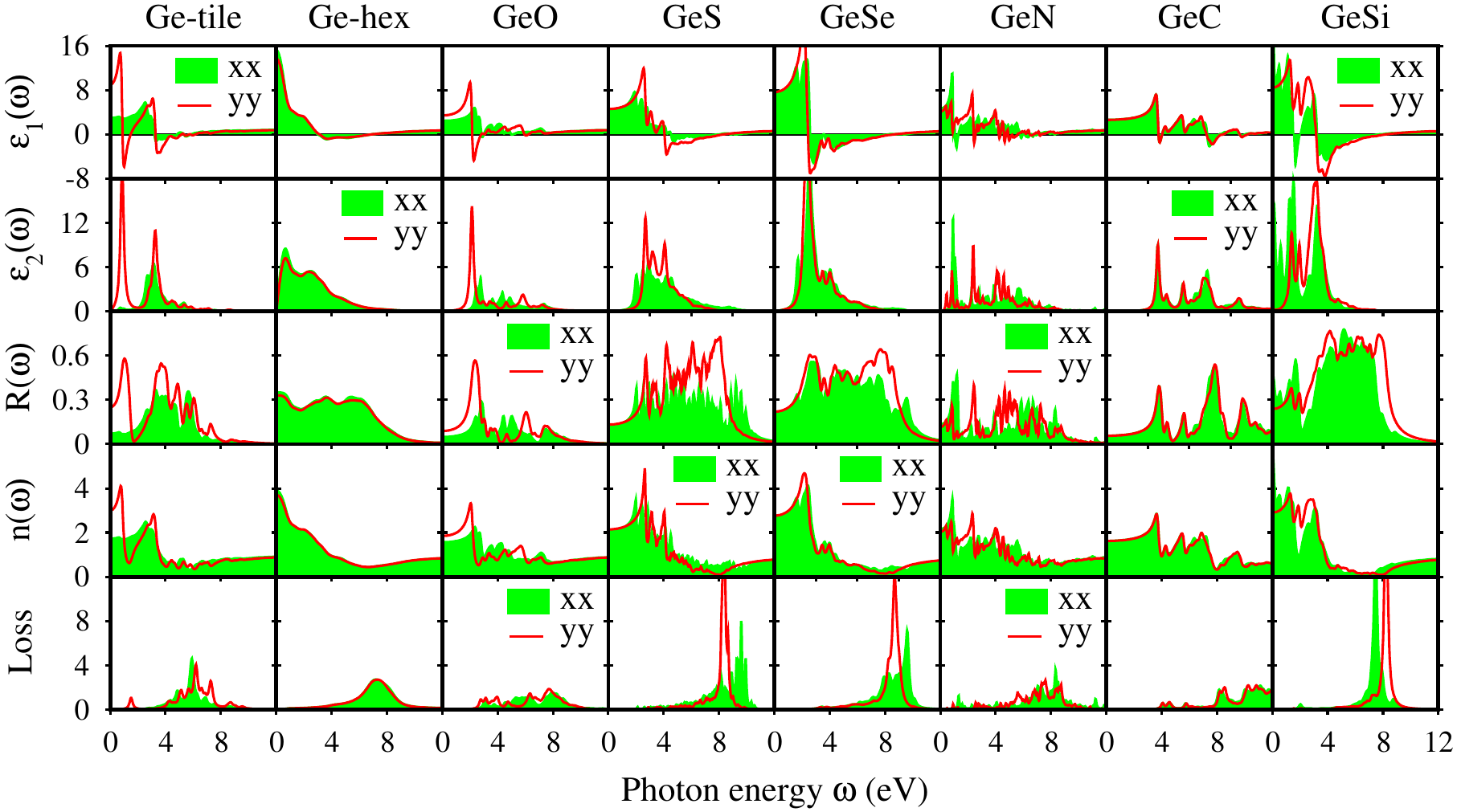}
\caption{\label{optic}
 Optical properties of our investigated 2D materials,
 including the real ($\varepsilon_1(\omega)$) and imaginary ($\varepsilon_2(\omega)$) 
 parts of the dielectric function, reflectivity ($\rm R(\omega)$),
 refractive index $\rm n(\omega)$, and energy loss function,
 calculated for the x (green shaded area) and y (red solid line) 
 polarizations of the incident light.
}
\end{figure*}

As it was mentioned in the section Method, we investigated the optical properties 
of our 2D materials in the framework of the Bethe-Salpeter approach.
The obtained optical properties including the real and imaginary parts of the dielectric function, 
reflectivity, refraction index, and the energy loss function are presented in Fig.~\ref{optic}.
Optical properties are calculated for two polarization of 
the incident light electric field along the x and y directions.
Comparing the $xx$ and $yy$ components of optical parameters
indicate that GeC and hexagonal germanene are well isotropic in 
the $xy$ plane in whole frequency range,
while other 2D systems exhibit clear in-plane anisotropies in 
the frequencies below 10\,eV,
being attributed to the anisotropic crystal structure of these materials. 
The observed in-plane anisotropy is more pronounced 
in GeO, GeS, and tile germanene.
It is interesting to see that the new invented structure of germanene
exhibit strong anisotropy in all optical parameters 
in low frequencies, below 2\,eV.
The strong anisotropy of the refractive index will likely lead to
birefringence behavior of tile germanene in low frequencies.
The results show high reflectivity of GeS, GeSe, GeSi, and hexagonal germanene
in a broad frequency range.
On the other hand, GeS, GeSe, and GeSi exhibit very low refractive index
around frequency of 8\,eV, in the UV region.

The first peak in the imaginary part of the dielectric function 
is expected to give the optical gap of semiconductors,
while the characteristic frequency of metals, corresponding to the collective excitations 
of valence electrons, known as plasma frequency, is given by those peaks of the loss function
which are located around nodes of the real part of the dielectric function. 
The optical gap and plasma frequencies of the systems
were determined and presented in table~\ref{static}.
The corresponding zero energy value of the real part of dielectric function,
known as static relative permittivity, 
is also extracted and given in this table.
It is seen that GeSi and hexagonal germanene shows 
the highest static relative permittivity among our studied systems. 
The largest optical gap is observed in GeC, 
while GeN has a small optical gap of about 0.45\,eV.
Since the calculated optical responses are obtained in the presence of
the attractive electron-hole interaction (excitonic effects),
well described in the Bethe-Salpeter approach,
the distance between the optical gap and the electric gap
is introduced as a measure of the exciton binding energy of the system.\cite{onida2002,yang2013}
In the framework of electronic structure theory, 
the electric gap is determined by the many body based GW scheme
and as it was mentioned before, the GW gaps are
expected to be very close to the obtained gaps
within the DFT1/2 scheme\cite{doumont2019} (table~\ref{bandgap}).
Hence, we calculated the exciton binding energy of our 2D systems
as the difference between their optical and DFT1/2 gaps and 
presented the results in table~\ref{static}.
We observe a very low exciton binding energy for GeC and GeSe,
which may indicate a very low carrier recombination rate 
in these systems after photo-excitations.
This observation encourages photocatalytic and photovoltaic
applications of 2D GeC and GeSe materials.

\begin{table}
\caption{\label{static}
 Obtained static value (at $\omega=0$) 
 of the real part of dielectric function,
 optical gap (eV), exciton binding energy $\Delta_X$ (eV), 
 and  plasma frequencies $\omega_p$ (eV) of the investigated 2D systems. 
}
\newcommand{\e[1]}{$\varepsilon^1_{#1#1}$}
\newcommand{\n[1]}{$n_{#1#1}$}
\begin{ruledtabular}
\begin{tabular}{lccccl}
        & \e[x]    & \e[y] & gap  & $\Delta_X$ & ~~~~$\omega_p$ \\
\hline                                          
GeO     &  2.42    &  3.33 & 2.15 &    1.07    & 2.85      \\
GeS     &  4.33    &  4.56 & 2.04 &    0.23    & 8.32      \\
GeSe    &  7.21    &  7.62 & 1.68 &    0.00    & 8.56      \\
GeN     &  4.61    &  4.09 & 0.45 &    0.84    & 1.33, 2.55 \\
GeC     &  2.63    &  2.63 & 3.50 &    0.00    & 4.05, 8.45 \\
GeSi    & 12.94\,~~&  8.59 & ---  &            & 2.02, 8.14 \\
Ge-tile &  2.93    &  8.70 & ---  &            & 1.48, 4.88 \\
Ge-hex  & 14.84\,~~& 13.49\,~~& ---  &         & ---       \\
\end{tabular}
\end{ruledtabular}
\end{table}

The last column in table~\ref{static} shows
the predicted plasma frequencies of our systems.
Although, plasma frequency is usually a characteristic feature
of metallic systems, as it is seen in the case of our 2D semiconductors,
semiconducting materials may also exhibit this kind of collective 
excitation of valence electrons.\cite{yu1996}
While most of the investigated 2D systems has a plasma frequency in the UV region,
we observe that plasma oscillations in the visible region may happen in 
the semiconducting GeO and GeN and metallic GeSi 2D materials.
The results show that, in the IR region, only metallic tile germanene and semiconducting GeN 
may exhibit plasmonic excitations.

\section{CONCLUSIONS}
In summary, we carried out a comprehensive first-principles study 
to investigate electronic and optical properties of eight Ge based 2D materials.
The calculated binding energies and phonon spectra indicate that GeO and GeN monolayers
stabilize in a zigzag structure, GeS and GeSe sheets prefer a puckered configuration,
and GeC occurs in a zero buckled (honeycomb) lattice.
In the case of germanene monolayer, we obtained a new structure, 
called as tile germanene, which is about 0.11\,eV/atom more stable
than the known hexagonal structure of germanene.
The obtained electronic structures suggest that tile germanene and GeSi
are very good semiconductors, while GeO, GeN, GeS, GeSe, and GeC display
a band gap in their electronic structure.
The novel DFT1/2 scheme was applied to obtain reliable and accurate band gaps.
The largest band gap is seen in GeC (3.50\,eV) while GeN
exhibits the smallest band gap (1.29\,eV) among these systems.
We used the vacuum level potential in the slab supercell as the energy reference
to determine the absolute band edges of the semiconducting systems.
The resulting valence and conduction band edges suggest potential application
of the GeO, GeS, and GeC monolayers as photocatalyst for water splitting.
The Bethe-Salpeter approach was used to compute various optical properties,
optical band gap, and plasma frequencies of the samples in
the presence of excitonic effects.
We observed a very low exciton binding energy in the GeC and GeSe sheets
which further encourage photocatalytic application of these 2D materials.

\section{ACKNOWLEDGMENTS}
This work was supported by the Vice Chancellor of
Isfahan University of Technology (IUT) in Research Affairs.

\bibliography{germanene}
\end{document}